\documentclass[12pt]{JHEP3}

\usepackage{epsfig}
\usepackage{subfigure}
\usepackage{wrapfig}

\usepackage{amsmath}
\usepackage{amssymb}
\usepackage{cite}
\usepackage{array}

\newcommand{\arcsinh}{{\rm arcsinh}\,}

\newcommand{\hyperg}{{\rm _2F_1}\,}

\newcommand{\beqs}{\begin{equation*}}
\newcommand{\beq}{\begin{equation}}

\newcommand{\eeqs}{\end{equation*}}
\newcommand{\eeq}{\end{equation}}

\newcommand{\beqas}{\begin{eqnarray*}}
\newcommand{\beqa}{\begin{eqnarray}}

\newcommand{\eeqas}{\end{eqnarray*}}
\newcommand{\eeqa}{\end{eqnarray}}

\newcommand{\eq}[2]{\begin{equation} #1 \label{#2} \end{equation}}

\newcommand{\eps}{\varepsilon}
\newcommand{\al}{\alpha}

\newcommand{\om}{\omega}

\newcommand{\si}{\sigma}

\newcommand{\blist}{\begin{itemize}}

\newcommand{\elist}{\end{itemize}}

\providecommand{\href}[2]{#2}

\title{Non-existence of a dilaton gravity action 
for the exact string black hole}

\author{D.\ Grumiller\\ Institut f\"ur Theoretische Physik, TU Wien, \\ Wiedner Hauptstr.\  8-10, A-1040 Wien, Austria\\ E-mail: \email{grumil@hep.itp.tuwien.ac.at}}

\author{D.V.\ Vassilevich\thanks{On leave from V.A.Fock Institute of Physics, St.Petersburg University, 198904 St.Petersburg, Russia}\\ Max-Planck-Institut f\"ur Mathematik in den Naturwissenschaften, \\ Inselstr. 22-26, D-04103 Leipzig, Germany\\ E-mail: \email{vassil@itp.uni-leipzig.de}}

\abstract{We prove that no local diffeomorphism invariant two-dimensional 
theory of the metric and the dilaton without higher derivatives
can describe the exact string black hole solution found a decade ago by
Dijkgraaf, Verlinde and Verlinde.
One of the key points in this proof is the concept of dilaton-shift 
invariance. We present and solve (classically) all dilaton-shift invariant 
theories of two-dimensional dilaton gravity.
Two such models, resembling the exact string black hole and generalizing the 
CGHS model, are discussed explicitly.}


\preprint{TU--02--21}

\begin{document}

\section{Introduction}

At the beginning of the last decade intense activity has been devoted to the
construction of conformal field theories representing strings propagating in
black hole backgrounds. One particularly successful example is the 
two-dimensional Witten black hole \cite{Mandal:1991tz,Elitzur:1991cb,
Witten:1991yr} resulting from a $SL(2,\mathbb{R})/U(1)$ gauged 
Wess-Zumino-Witten model.
Detached from its stringy origin it has inspired the influential 
paper of Callan, Giddings, Harvey and Strominger (CGHS) \cite{Callan:1992rs} 
which rekindled the interest in two-dimensional (dilaton) gravity in the early
1990's (for a recent review cf. \cite{Grumiller:2002nm}\footnote{
Several relevant papers on string gravity in two dimensions
\cite{Chamseddine:1989tu,Chamseddine:1989yz,Chamseddine:1990wn,
Chamseddine:1991hr}
were omitted in the
printed version of \cite{Grumiller:2002nm}.}).

However, in Witten's original work the metric and dilaton satisfy the 
corresponding $\si$-model conformal invariance conditions only to lowest 
order. By conformal field theory methods (which are somewhat indirect in this
context) Dijkgraaf, Verlinde and Verlinde 
presented a solution for the metric and the dilaton, the exact string 
black hole (ESBH) \cite{Dijkgraaf:1992ba}, which supposedly solves the problem 
non-perturbatively. Indeed, it has been shown that the ESBH is consistent with 
$\si$-model conformal invariance up to three loops in the bosonic case 
\cite{Tseytlin:1991ht} and up to four loops in the supersymmetric one 
\cite{Jack:1993mk}. For further historical and technical details cf.\ e.g.\ 
\cite{Becker:1994vd} and references therein.

What is still lacking is a non-perturbative effective action. 
In this paper we address this issue and try to construct a dilaton model 
reproducing the ESBH solution.
We fail, but take revenge and show in turn that, indeed, such a 
construction is impossible with the given assumptions (local 
Lorentz-invariance, locality, local diffeomorphism invariance, absence of
propagating degrees of freedom and $D=2$).  
Since the dimensionality, local Lorentz invariance and local diffeomorphism 
invariance should be kept by all means (after all, a description in terms of a 
two-dimensional metric is desired) this seems to imply that either 
locality must be violated or higher derivative interactions must 
appear in the action thus leading to propagating modes in this model. 
%NEWDANIEL
We should stress that in our approach we do not exploit the fact that the ESBH 
follows from an exact conformal field theory (CFT). A more extensive 
discussion on these points can be found in section \ref{sec:5}.

As by-products we present two models which are close to the ESBH solution and
justify a study on their own. 

This work is organized as follows:

In section \ref{sec:2} we review briefly the ESBH solution and fix most of 
our notations. Section \ref{sec:3} summarizes generalized dilaton gravity 
in the first order formalism and introduces the important concept of 
dilaton-shift invariance, a property which must be shared by any model
describing the ESBH solution. All classical solutions are obtained.
In section \ref{sec:4} we prove that no such action compatible with the ESBH
exists. Section \ref{sec:5} concludes this work. 

Supplementary material can be found in the two appendices: Appendix \ref{app:A}
investigates the most general form of a dilaton-shift invariant model. In 
appendix \ref{app:B} two promising toy-models are discussed, resembling the
ESBH in many relevant aspects.
 
\section{Exact string black hole}\label{sec:2}

In the notation of \cite{Kazakov:2001pj} the line element of the ESBH 
discovered by Dijkgraaf, Verlinde and Verlinde \cite{Dijkgraaf:1992ba} is 
given by
\eq{
(ds)^2=(dx)^2+f^2(x)(d\tau)^2\,,
}{eq:dvv1}
with
\eq{
f(x)=\frac{\tanh{(bx)}}{\sqrt{1-p\tanh^2{(bx)}}}\,.
}{eq:dvv2}
Physical scales are adjusted by the parameter $b\neq 0$ which has dimension of
inverse length. The corresponding expression for the dilaton,  
\eq{
\phi=\phi_0-\ln{\cosh{(bx)}}-\frac{1}{4}\ln{(1-p\tanh^2{(bx)})}\,,
}{eq:dvv3}
contains an integration constant $\phi_0$.
Additionally, there are the following relations between constants, 
string-coupling $\al\prime$, level $k$ and dimension $D$ of string target 
space:
\eq{
\alpha\prime b^2=\frac{1}{k-2}\,, \hspace{0.5cm}
p:=\frac{2}{k}=\frac{2\alpha\prime b^2}{1+2\alpha\prime b^2}\,, \hspace{0.5cm} 
D-26+6\alpha\prime b^2=0\,.
}{eq:dvv4}
For $D=2$ one obtains $p=8/9$, but like in the original work 
\cite{Dijkgraaf:1992ba} we will treat general values of $p\in(0;1)$.

In the present work exclusively the Minkowskian version of (\ref{eq:dvv1})
\eq{
(ds)^2=f^2(x)(d\tau)^2-(dx)^2\,,
}{eq:dvv5}
will be needed. With the definitions
\eq{
\sqrt{1-p}f(x)dx =: dr\,, \hspace{0.5cm}(1-p)f^2(x(r)) =: \xi(r)\,,
\hspace{0.5cm}\frac{d\tau}{\sqrt{1-p}} =: dt\,,
}{eq:dvv6}
the line element can be presented, for instance, in Schwarzschild gauge
$(ds)^2=\xi(r)(dt)^2-\xi^{-1}(r)(dr)^2$ or with $du:=dt-\xi^{-1}(r)dr$ in 
Eddington-Finkelstein gauge
\begin{equation}
(ds)^2=2du\otimes dr+\xi(r)(du)^2\,, 
\label{dvvEFg}
\end{equation}
identifying $\xi(r)$ as the Killing-norm. Since we are going to suppress the
wedge symbol $\wedge$ subsequently we keep the symmetrized direct product 
symbol $\otimes$ to avoid confusion\footnote{The notation $(du)^2$ means 
$du\otimes du$, but this is rather obvious.}.
The curvature scalar for the metric (\ref{eq:dvv5}) reads
\begin{equation}
R=2f(x)^{-1}\partial_x^2 f(x)
=\frac{2b^2(3p-2-p \tanh^2(bx))}{\cosh^2(bx)(1-p\,\tanh^2(bx))^2} \,.
\label{dvvR}
\end{equation}
The maximally extended space-time of this geometry has been studied by
Perry and Teo \cite{Perry:1993ry} and by Yi \cite{Yi:1993gh}.

For the rest of this work we will assume $p\in(0;1)$ and 
consider the limits $p\to 0$ and $p\to 1$ separately: for $p=0$ one recovers
the CGHS model; for $p=1$ the Jackiw-Teitelboim (JT) model 
\cite{Barbashov:1979bm,D'Hoker:1982er,Teitelboim:1983ux,D'Hoker:1983is,D'Hoker:1983ef} 
is obtained. Both limits exhibit singular features: 
for all $p\in(0;1)$ the solution is regular 
globally, asymptotically flat and exactly one Killing-horizon exists. However, 
for $p=0$ a singularity (screened by a horizon) appears and for $p=1$ 
space-time fails to be asymptotically flat.

Winding/momentum mode duality implies the existence of a dual solution which 
can be acquired most easily by replacing $bx\to bx+i\pi/2$, entailing in all 
formulae the substitutions 
\eq{
\sinh\to i\cosh\,,\hspace{0.5cm}\cosh\to i\sinh\,.
}{dvv:dual}

Note that the integration constant $\phi_0$ enters only the dilaton
field, but not the metric. Therefore, a symmetry property exists which
proves very important: constant shift of the dilaton $\phi$ maps
a solution to another one of the same model.
%%%%%%%%
\section{Generalized dilaton theories}\label{sec:3}
\subsection{First and second order actions}
We start with the first order action for dilaton gravity in two 
%NEWDANIEL
dimensions \cite{Schaller:1994es},
\begin{equation}
L^{(1)}=\int_{\mathcal{M}_2} \left[
X_a (De)^a +Xd\omega +\epsilon \mathcal{V} (X_aX^a, X) \right]\,,
\label{eq:dvva1}
\end{equation}
where $X$ is the dilaton field, $e^a$ is the zweibein one-form,
$\epsilon$ is the volume two-form. The one-form $\omega$ represents the 
spin-connection $\om^a{}_b=\eps^a{}_b\om$ with $\eps_{ab}$ being the totally
antisymmetric Levi-Civit{\'a} symbol. The action (\ref{eq:dvva1}) 
depends on two auxiliary fields $X^a$. It is a special case of a 
Poisson-$\si$ model \cite{Ikeda:1993aj,Ikeda:1994fh,Schaller:1994es}
with a three dimensional target space\footnote{When referring to 
``target-space'' from now on we mean the Poisson manifold, not the 
target-space of string theory.} the
coordinates of which are $X,X^a$. In light-cone coordinates 
($\eta_{+-}=1=\eta_{-+}$, $\eta_{++}=0=\eta_{--}$) the first (``torsion'') 
term of (\ref{eq:dvva1}) is given by
\begin{equation}
X_a(De)^a = \eta_{ab}X^b(De)^a =X^+(d-\omega)e^- +
X^-(d+\omega)e^+\,.\label{dvvXDe}
\end{equation}
The function $\mathcal{V}$ is an arbitrary potential depending solely on
Lorentz invariant combinations of the target space coordinates, namely
$X$ and $X^+X^-$.

In string physics the CGHS \cite{Callan:1992rs} and JT \cite{Barbashov:1979bm,D'Hoker:1982er,Teitelboim:1983ux,D'Hoker:1983is,D'Hoker:1983ef}
models play a special role. They correspond to specific choices
of the potential $\mathcal{V}$ ($b\neq 0$ is a constant):
\begin{eqnarray}
&&\mathcal{V}_{CGHS}= -\frac{X^+X^-}{X} - 2b^2X\,,\label{dvv:VCGHS}\\
&&\mathcal{V}_{JT}= -b^2X\,.\label{dvv:VJT}
\end{eqnarray}

The auxiliary fields can be eliminated by means of their (algebraic)
equations of motion (EOMs). The action (\ref{eq:dvva1}) is equivalent to
the following second order action
\begin{equation}
L^{(2)}=\int_{\mathcal{M}_2} \left[
\frac{XR}2 +\mathcal{V}(-(\nabla X)^2,X)\right]\sqrt{-g}d^2x\,.
\label{dvva2}
\end{equation}
For supplementary information on dilaton gravity in two dimensions the 
recent review \cite{Grumiller:2002nm} may be consulted.

In principle, $X$ may be an arbitrary local function of the dilaton
$\phi $ appearing in the ESBH solution (\ref{eq:dvv3}).
Also, ${\mathcal{V}}$ is an arbitrary function of two variables.
The model looks too general to be handled effectively. Therefore,
our next step is to find restrictions on $X$ and $\mathcal{V}$
which follow from the dilaton-shift invariance $\phi\to\phi+\phi_0$, 
$\phi_0\in\mathbb{R}$ discussed at the end of the previous section.  
There are two types of dilaton actions which respect this symmetry. 

The first one 
\begin{equation}
X=\phi \,,\qquad \mathcal{V}(X_aX^a,X)=
\tilde U(X^+X^-)\,, \label{dvvXtU}
\end{equation}
contains an arbitrary function $\tilde U$. However, the EOM for the field $X$ 
requires $R=0$ identically which contradicts (\ref{dvvR}) and rules out this 
(rather trivial) variant.

The second possibility is
\begin{equation}
X=\exp (-2\alpha \phi )\,,\qquad 
\mathcal{V}(X_aX^a,X)= XU\left( \frac{X^+X^-}{X^2} \right)\,,
\label{dvvXU}
\end{equation}
where $U$ is an arbitrary function of one variable and $\alpha$ an
arbitrary constant. Under the transformation $\phi \to \phi + {\rm const.}$
the action (\ref{dvva2}) with (\ref{dvvXU}) changes by a multiplicative
constant leaving the EOMs invariant. Note, that the CGHS model
 (\ref{dvv:VCGHS}) and the JT model (\ref{dvv:VJT}) belong to this
category.

We remark that the condition of dilaton-shift invariance of the action
requires homogeneity of the potential $\mathcal{V}$ under rescalings
of the dilaton field. This naturally reduces the initial freedom in
the problem (arbitrary function of two variables) to an arbitrary
function of a single scale-invariant variable.

In appendix \ref{app:A} we prove that no further dilaton-shift invariant
actions of type (\ref{eq:dvva1}) resp.\ (\ref{dvva2}) exist. Thus, it must be
possible to reconstruct the ESBH solution from (\ref{eq:dvva1}) with  
(\ref{dvvXU}) or no action of type (\ref{eq:dvva1}) exists which produces it.

%%%%%%%%%%%%%%
\subsection{All classical solutions of the dilaton-shift invariant 
models}\label{solutions}

All non-trivial dilaton-shift invariant first order actions follow from 
(\ref{eq:dvva1}) and 
(\ref{dvvXU}):
\begin{equation}
L=\int_{\mathcal{M}_2} \left[ X_a(De)^a +Xd\omega +\epsilon X\, 
U\left(Z\right) \right]\,, \label{dvva3}
\end{equation}
where
\begin{equation}
Z:=\frac{X^+X^-}{X^2} \,.\label{dvvZ}
\end{equation}
The EOMs derived from such an action read
\begin{eqnarray}
&&dX+X^-e^+-X^+e^-=0\,,\label{dvveo}\\
&&(d\pm \omega )X^\pm \mp e^\pm XU(Z)=0\,,\label{dvvee}\\
&&d\omega +\epsilon \left( U(Z) -2ZU'(Z) \right) =0\,,
\label{dvveX}\\
&&(d\pm \omega )e^\pm +\epsilon \frac{X^\pm}{X} U'(Z)=0\,.
\label{dvveXpm}
\end{eqnarray}

Integrability of this model can be deduced from general Poisson-$\si$
model 
%NEWDANIEL
arguments\footnote{The local and global solutions of generic 
dilaton gravity in $D=2$ have been discussed in refs.\ \cite{Klosch:1996fi,
Klosch:1996qv,Klosch:1997md}.} and is closely related to the existence of a 
(Casimir) function $\mathcal{C}$ depending on the target-space coordinates 
which is absolutely conserved:
\begin{equation}
d\mathcal{C}(X,X^+X^-)=0 \,.\label{dvvCcon}
\end{equation}
To find this quantity we multiply the first of the equations
(\ref{dvvee}) by $X^-$ and add it to the second one
multiplied by $X^+$. Then we employ (\ref{dvveo}) to obtain
\begin{equation}
d(X^+X^-)+U(Z) XdX=0 \,.\label{dvvdX}
\end{equation}
This equation is equivalent to the conservation law (\ref{dvvCcon})
with $\mathcal{C}$ given by
\begin{equation}
\mathcal{C}=X\exp (W(Z))\,,\qquad
W(Z):=\int^Z \frac{dY}{U(Y)+2Y} \,.
\label{dvvCW}
\end{equation}
On each solution $\mathcal{C}$ is a constant, $\mathcal{C}=\mathcal{C}_0$.
Therefore, eq.\ (\ref{dvvCW}) permits to express $Z$ in terms of $X$,
\eq{
Z = W^{-1} \left(\ln{\frac{\mathcal{C}_0}{X}}\right)\,.
}{eq:dvv47}

Next we assume $X^+\ne 0$ in a given patch\footnote{
If $X^+=0$, we can repeat the subsequent calculations with
the index $+$ replaced by $-$ everywhere. If both $X^+$ and
$X^-$ are zero in a patch, eq.\ (\ref{dvveo}) yields
$X=\rm const.$ in this patch, which is not the case (cf.\ (\ref{eq:dvv3})).}
and define a new one-form $f$,
\begin{equation}
f=e^+/X^+ \,.\label{dvvf}
\end{equation}
Eq.\ (\ref{dvveo}) establishes
\begin{equation}
e^-=\frac{dX}{X^+} +X^-f \,,\qquad
\epsilon =e^+\wedge e^-=-dX\wedge f \,.\label{dvveep}
\end{equation}
Eqs.\ (\ref{dvvee}) and (\ref{dvveXpm}) yield
\begin{eqnarray}
&&\omega = XU(Z)f-dX^+/X^+ \,,\label{dvvom}\\
&&df=\frac{U'(Z)}{X} dX\wedge f\,.
\label{eq:dvv45}
\end{eqnarray}
Thus eq.\ (\ref{eq:dvv45}) can be integrated,
\eq{
f=\tilde{f}\exp{\int^XV(X')dX'}=:\tilde{f}I(X)\,,
}{eq:dvv49}
where
\begin{equation}
V(X):=\frac{U'\left( W^{-1}(\ln (\mathcal{C}_0/X))\right)}X
\label{dvvV}
\end{equation}
is a given function of $X$ for each particular model. The ``integration
constant'' $\tilde f$ is now a one-form which should satisfy
$d\tilde f=0$. Therefore, the equality $\tilde f=d\tilde u$ is valid
at least locally. $\tilde u$ can be
used as one of the coordinates. It is convenient to choose
$X$ as the second one. The line-element $(ds)^2=2e^-\otimes e^+$ emerges
as
\eq{
(ds)^2= I(X)\left[2dX\otimes d\tilde u+2(d\tilde u)^2 X^2 
W^{-1}\left(\ln{\frac{\mathcal{C}_0}{X}}\right)I(X)
\right]\,.
}{eq:dvv52}

Lower limits of the integrals in (\ref{dvvCW}) and 
(\ref{eq:dvv49}) are arbitrary. Shifts of them
may be absorbed into a re-definition of $\mathcal{C}$ and into
a rescaling of the coordinate $\tilde u$, respectively. We shall exploit
this freedom in the next section.
%%%%%%%%%%%%%%%%%%%%%%%%
\section{(An attempt of) construction of the potential $\boldsymbol{U(Z)}$}\label{sec:4}
%%%%%%

As a preamble why this construction could be possible we consider the limits
$p\to 0$ and $p\to 1$. In the former case the CGHS model is recovered with
a linear potential $U_{CGHS}(Z)=-2b^2-Z$ (cf.\ eq.\ (\ref{dvv:VCGHS})). The 
latter limit induces the JT model with a constant potential $U_{JT}(Z)=-b^2$ 
(cf.\ eq.\ (\ref{dvv:VJT})).

Because of our ability to describe both limits with the desired class of 
models it seems plausible that an interpolating theory of the same structure 
describing the ESBH for all values of $p$ could exist. However, as plausible 
as it may be, it is not true unfortunately, as will be proved in this section.

\subsection{Restrictions on $U(Z)$ following from the line element}
To construct a potential $U(Z)$ which reproduces the ESBH 
solutions one has first to convert the metric
(\ref{dvvEFg}) to a form admitting a comparison with
(\ref{eq:dvv52}). To this end we change the independent variable $r$
to $X$, where $X$ is the exponentiated dilaton defined by (\ref{dvvXU}),
(\ref{eq:dvv3}),
\begin{equation}
(ds)^2=2g_{uX}du \otimes dX+g_{uu}(du)^2\,.
\label{dvvdsuX}
\end{equation}
Obviously, the simple relations
\begin{equation}
g_{uX}=\frac{dr}{dX} =\frac{dr/dx}{dX/dx} \label{dvvguX}
\end{equation}
must hold. One readily obtains
\begin{equation}
g_{uX}=X^{(1-\alpha)/\alpha} \sqrt{1-p}
\left[ \alpha b e^{-2\phi_0} \left(
2(1-p)\cosh^2 (bx) +p \right) \right]^{-1} \,.
\label{dvvguX1}
\end{equation}
By using the identity\footnote{Since $X\in\mathbb{R}_0^+$,
$\phi_0\in\mathbb{R}$, $p\in(0;1)$ and $\cosh(x)\in\mathbb{R}^+$ for 
$x\in\mathbb{R}$ there are no ambiguities involved.}
\begin{equation}
\cosh^2(bx)=\frac{-p+\sqrt{ 4(1-p) X^{2/\alpha} e^{4\phi_0} +p^2}}{
2(1-p)}\label{dvvcosh2}\,,
\end{equation}
eq.\ (\ref{dvvguX1}) can be rewritten in terms of $X$ only,
\begin{equation}
g_{uX}=X^{(1-\alpha)/\alpha} \sqrt{1-p}
\left[ \alpha b e^{-2\phi_0} 
\sqrt{ 4(1-p) X^{2/\alpha} e^{4\phi_0} +p^2} \right]^{-1}\,.
\label{dvvguX2}
\end{equation}
Constant rescaling of $u$ is the only residual gauge freedom left
in the line element (\ref{dvvdsuX}) as discussed at the end of the previous 
section. Consequently, $du$ and $d\tilde u$ must be equal up to a constant 
$du=\mu d\tilde u$. From (\ref{eq:dvv52}) it can be extracted promptly that
this ambiguity can be put into $I(X)$, 
\begin{equation}
I(X)=\mu g_{uX} \,.\label{dvvmu}
\end{equation}
Clearly, $V(X)$ does not depend on the scale factor $\mu$:
\begin{equation}
V(X)=\frac{\partial}{\partial X} \ln I(X)=\frac 1{\alpha X}
\left[ 1-\alpha -\left( 1 +\frac{p^2 X^{-2/\alpha}}{4(1-p)e^{4\phi_0}}
\right)^{-1} \right]
\label{dvvVX}
\end{equation}

Now we have to re-express $g_{uu}=\xi$ in terms of $X$:
\begin{equation}
g_{uu}=1 - \frac{2}{p(1+w)}\,,\label{dvvguu}
\end{equation}
where
\begin{equation}
w:=\sqrt{ \frac{4(1-p)X^{2/\alpha}e^{4\phi_0}}{p^2} +1}\,. 
\label{dvvw}
\end{equation}
Because $w$ ranges from $1$ to $+\infty$ the Killing-norm (\ref{dvvguu}) has 
no poles, but one zero for each $p\in(0;1)$. 
In complete analogy to the previous calculations the substitutions 
(\ref{dvv:dual}) imply for the dual Killing-norm
\eq{
g_{uu}^{\rm dual} = \xi^{\rm dual} = 1 - \frac{2}{p(1-w)}\,.
}{dvv:dualkilling}
It exhibits no zeros, but one pole for all $p\in(0;1)$. Formally, duality in 
this scenario can be interpreted as a branch cut ambiguity: if $w$ as defined 
in (\ref{dvvw}) is replaced by $-w$ (i.e. if we go the the second branch of the
square-root) we perform a duality transformation. This observation will be 
exploited further in appendix \ref{app:B}.

From the line element (\ref{eq:dvv52}) $Z(X)=W^{-1}(\ln (\mathcal{C}_0/X))$ can
be identified: 
\begin{equation}
Z(X)=\frac{2\alpha^2 b^2 w^2((w+1)p -2)}{p(w^2-1)(1+w)} \,.
\label{dvvZX}
\end{equation}
Note, that there is no dependence on the rescaling
$\mu$. $U'(X)$ is immediately read off from (\ref{dvvV}) and
(\ref{dvvVX}):
\begin{equation}
U'(X)=\frac 1\alpha \left( \frac 1{w^2} -\alpha \right)\,,
\label{dvvUX}
\end{equation}
where $U'(X)$ is just a short-hand notation for
$U'(W^{-1}(\ln (\mathcal{C}_0/X)))$.

That is already enough to define $U(Z)$. From (\ref{dvvZX}) $w$ can be 
expressed as a function of $Z$ by solving a cubic equation. Together with 
eq.\ (\ref{dvvUX}) this gives $U'$ as a function of $Z$, an ordinary
differential equation which can be integrated thus determining $U(Z)$
up to several free parameters (the detailed construction is pursued 
in appendix \ref{app:B.3}). Then one may apply the results of 
sec.\ \ref{solutions} to construct classical solutions for all
potentials $U'$ obtained in this way and compare with the
ESBH solution. In principle, this procedure allows
either to fix the undetermined parameters in $U$ or, if suitable
parameters do not exist, to demonstrate that there is no dilaton
gravity model describing the ESBH. However, it leads
to considerable technical difficulties already at the first
steps. %(cf.\ appendix \ref{app:B}).
We shall choose another way.
%%%%
\subsection{Restrictions on $U(Z)$ following from the conserved quantity}
So far we have not taken into account the fact that the potential $U(Z)$
defines the functional form of the conserved quantity $\mathcal{C}$
and, through the eqs.\ (\ref{dvvCW}) and (\ref{eq:dvv47}),
dependence of $Z$ on $X$ for each solution of the EOMs. To simplify the 
analysis let us introduce two new functions depending on $U$:
\eq{
U(Z)=:-2Z+y(Z)\,,
}{eq:dvvnew1}
and
\eq{
y(Z)=:\frac{k(Z)}{k'(Z)}\,.
}{eq:dvvnew1a}
Eqs. (\ref{dvvCW}), (\ref{eq:dvvnew1a}) yield the simple result
\eq{
k(Z)= \frac{{\cal C}_0}{X}\,,
}{eq:dvvnew2}
having absorbed a multiplicative constant of $k(Z)$ into ${\cal C}_0$. 
We still possess this freedom since the lower limit in the integral
in (\ref{dvvCW}) has not been fixed.
Note that from now on we are working with a selected solution.
Therefore, $U$, $k$, $y$, $Z$, and $X$ are functions of a single
coordinate. This makes all derivatives of these functions with
respect to each other well defined. In particular, we need
\begin{equation}
\frac{dy}{dk}=\frac{dy/dZ}{dk/dZ}=
\frac yk \left[ \frac 1{\alpha w^2} +1 \right]\,,
\label{dvvdydk}
\end{equation}
having inserted the definitions (\ref{eq:dvvnew1}) and (\ref{eq:dvvnew1a})
together with (\ref{dvvUX}). Because of eq.\ (\ref{eq:dvvnew2})
$w$ can be considered as a function of $k$. Therefore, (\ref{dvvdydk})
is an ordinary differential equation which can be solved straightforwardly,
\begin{equation}
y=y_0 k \left( k^{2/\alpha} +B\right)^{\alpha/2}
\,,\qquad B:= \frac{4(1-p)\mathcal{C}_0^{2/\alpha}e^{4\phi_0}}{p^2}\,,
\label{dvvyy0}
\end{equation}
with $y_0$ as integration constant. Furthermore, eq.\ (\ref{eq:dvvnew1a})
provides
\begin{equation}
\frac {dk}{dZ} =y_0^{-1} \left( k^{2/\alpha} +B\right)^{-\alpha/2} 
\,,\label{dvvdkdZ}
\end{equation}
or, equivalently,
\begin{equation}
\frac{dZ}{dX} =-y_0 \mathcal{C}_0^2 X^{-3} w^\alpha \,.
\label{dvvdZdX}
\end{equation}
%%%%
\subsection{The inconsistency}\label{sec:4.3}
By comparing (\ref{dvvZX}) and (\ref{dvvdZdX}) one sees that it is
unlikely that the model can be made consistent by a suitable choice
of the integration constants. The simplest way to demonstrate the 
incompatibility of these equations is to consider $p\to 0$:
\begin{equation}
(pw)=2e^{2\phi_0}X^{1/\alpha}+\mathcal{O}(p^2)
\label{dvvpw}
\end{equation}
Plugging this limit into eq.\ (\ref{dvvZX}) yields
\begin{equation}
Z(X)=2\alpha^2b^2 \left[ 1-X^{-1/\alpha}e^{-2\phi_0} +
\frac p2 X^{-2/\alpha} e^{-4\phi_0} \right] +\mathcal{O}(p^2)\,,
\label{dvvZp0}
\end{equation}
and consequently
\begin{equation}
\frac{dZ}{dX}=2\alpha b^2 \left[ X^{-1-\frac{1}{\alpha}}e^{-2\phi_0}
-p X^{-1-\frac{2}{\alpha}} e^{-4\phi_0} \right]  +\mathcal{O}(p^2)
\,.\label{dvvp1}
\end{equation}
On the other hand, (\ref{dvvdZdX}) demands
\begin{equation}
\frac{dZ}{dX}=- \tilde y_0 \mathcal{C}_0^2 \left( 2e^{2\phi_0} \right)^{\alpha}
X^{-2} +\mathcal{O}(p^2) \,,\label{dvvp2}
\end{equation}
where $\tilde y_0 =y_0 p^{-\alpha}$. For any finite $\alpha$ eq.\ 
(\ref{dvvp2}) contains a single power of $X$, while (\ref{dvvp1})
depends on two different powers. We conclude, that (\ref{dvvp1})
and (\ref{dvvp2}) are mutually incompatible\footnote{The limit 
$p\to 0$ is not necessary. One can reach the
same conclusion for arbitrary $p$ at the expense of somewhat more
involved calculations (cf.\ appendix \ref{app:B}).
If $p=0$ exactly (CGHS model), there is no contradiction. One obtains 
$\alpha=1$.
%NEWDANIEL
Expanding around $p=1$ yields essentially the same result: if $p=1$ exactly 
(JT model) no inconsistency arises for $\al=1$ (and proper adjustment of other
constants); however, for slightly smaller values of $p$ eq.\ (\ref{dvvZX}) 
yields $Z=Z_0+Z_1 X^{-2/\al}$ for any value of $\al\neq 0$ while 
(\ref{dvvdZdX}) provides $Z=Z_0+Z_1X^{-2}+Z_2X^{-2+2/\al}$ (with $Z_i\neq 0$)
for $\al\neq 0,1$ and gives rise to a logarithmic contribution for $\al=1$ 
showing again the incompatibility.
}. 

{\em Therefore, no dilaton gravity model (\ref{dvva2}) can generate the ESBH 
solution.}
%%%%%%%%%%
\section{Conclusions and outlook}\label{sec:5}

We start our conclusions with some generally accepted statements underpinning 
our line of reasoning: Each local diffeomorphism invariant two-dimensional 
theory of the metric and the dilaton without propagating degrees of freedom 
is generalized dilaton gravity (\ref{dvva2}). Locality and absence of 
propagating degrees of freedom excludes higher derivative terms\footnote{
The action (\ref{dvva2}) contains arbitrary powers of {\it
first} derivatives of $X$ (velocities) and is, therefore, local.}
as well as higher powers of the curvature\footnote{Powers of $R$
can appear after the dilaton $X$ has been eliminated by means of
its EOMs. Presence of higher curvature terms in the perturbative
string $\beta$-functions 
\cite{Tseytlin:1991ht,Metsaev:1987bc,Foakes:1987ij,Graham:1987ep,
Jack:1989vp} does not necessarily
imply that higher powers of $R$ should also appear in the action.
The $\beta$-functions are some (unknown) combinations of the EOMs rather than 
the EOMs themselves.}. 
% Diffeomorphism invariance restricts further the possible form of the potential ${\cal V}$. 
Equivalence to the first order formulation (\ref{eq:dvva1}) allows to apply 
the powerful tools available for Poisson-$\si$ models.

In this paper we have demonstrated that no such theory describes the exact 
string black hole solution found by Dijkgraaf, Verlinde and Verlinde 
\cite{Dijkgraaf:1992ba}. In the proof the property which we have called
``dilaton-shift invariance'' played a pivotal role. We discussed all such 
non-trivial models and found that none of them generates the correct solution.
This confirms the observation \cite{Grumiller:2002dm} that string gravities
occupy a special place among $2D$ gravity models.

As by-products we presented two dilaton-shift invariant toy-models which mimic
most of the desired features of the exact string black hole (and which could
be improved further by tinkering with certain fixing conditions for the two
essential constants involved).

What can be learned from these results?

First of all, the dilaton-shift invariant models are interesting on their own 
and deserve separate studies. While the CGHS model approximates the ESBH
in the weak coupling limit, and the JT model works in the strong coupling
regime, other dilaton-shift invariant models can be regarded as approximate
models which are uniformly good (or bad) for the whole range of $p$.
Some candidate models are presented in Appendix \ref{app:B}. We expect that by 
modifying the requirements listed at the beginning of Appendix \ref{app:B.1}
one can achieve a better agreement with ESBH. The gain from having such
approximate models is rather 
%NEWDANIEL
obvious: they are classically
integrable and presumably even the path integral quantization can be performed
exactly, as for other dilaton models in $2D$ \cite{Kummer:1997hy}. Another
interesting problem is to trace the action of string dualities on the
potential $U(Z)$.

Finally, the main result of this work has to be elucidated.
The most probable explanation of the non-existence of an effective action
of type (\ref{dvva2}) reproducing the exact string 
black hole solution is that 
there are indeed some higher order curvature or higher derivative
terms as suggested by perturbative results \cite{Tseytlin:1991ht}. 
If higher derivatives appear polynomially, this means that there are
some new degrees of freedom of the low energy string in the 
dilaton-graviton sector. If higher derivatives enter non-polynomially,
e.g. in a form of the inverse Laplacian, the action becomes non-local.
%NEWDANIEL
%Such a situation contradicts basic principles of string physics and is not likely. 
However, in many cases non-localities may be removed 
at the expense of introducing new fields. Therefore, the whole effect
may be similar to the case of polynomial higher derivative terms.
The presence of such terms would require a modification of the standard 
boundary term invalidating previous calculations of the ADM mass for the ESBH,
but also explaining why the approaches of refs.\
\cite{Witten:1991yr,Gibbons:1992rh,Nappi:1992as,Perry:1993ry,Liebl:1997ti,
Kazakov:2001pj} yield different values of 
the ADM mass\footnote{For a recent discussion on the ADM calculations for the 
ESBH cf.\ sec.\ 3 of ref.\ \cite{Kazakov:2001pj}.}.
%NEWDANIEL
Moreover, we have so far ignored the CFT aspects: the fact 
that the ESBH follows from an exact CFT and the 
corresponding $SL(2,\mathbb{R})/U(1)$ coset structure could help in 
determining these additional terms in the action. %\footnote{In this context it will be helpful to consider conformal transformations with a conformal factor $\Om(X)$ (because of the presence of one Killing field the conformal factor should only depend on the dilaton) which in the PSM formalism can be interpreted as target-space diffeomorphisms: $X=\tilde{X}$, $X^\pm=\tilde{X}^\pm\Om^{-1}$, $e^\pm=\tilde{e}^\pm\Om$, $\om=\tilde{\om}+(X^+e^-+X^-e^+)\Om'/\Om$. Under such a transformation the action (\ref{eq:dvva1}) is mapped onto itself (putting a tilde on top of all quantities) with a new potential $\tilde{{\cal V}}(\tilde{X}^+\tilde{X}^-,\tilde{X}) = \Om^2{\cal V}(\tilde{X}^+\tilde{X}^-\Om^{-2},\tilde{X})-2\tilde{X}^+\tilde{X}^-\Om'/\Om$. It is obvious that in this way a dilaton-shift invariant model will be mapped onto a model without this invariance (unless the function $U$ in ref.\ (\ref{dvva3}) is linear and homogeneous in $Z$ and the conformal factor is homogeneous in $X$). This is maybe not so surprising considering that we assumed invariance of curvature and torsion under dilaton shifts in appendix \ref{app:A}.  When adding conformal transformations to the mix one could weaken these requirements and impose instead dilaton-shift invariance up to conformal transformations. A plausible strategy would be to take the solution providing the correct dilaton discussed in appendix \ref{app:B.1} and then perform a conformal transformation where the conformal factor is chosen such that the ESBH line element is produced. However, the resulting transformed potential $\tilde{{\cal V}}$ will in general contain a ${\cal C}_0$ dependence wreaking havoc with this approach, because ${\cal C}_0$ must be a parameter labelling the solutions rather than being part of the action.}.
Of course, also the pessimistic variant, the failure of the ESBH
being the result of an effective $\si$-model action, cannot be ruled out.
A more exciting explanation could be that the Poincar{\'e} algebra
is being deformed thus requiring a different form of the dilaton 
action\footnote{Deformations of dilaton gravities in $2D$ were recently 
considered in ref.\ \cite{Mignemi:2002hd}.}.

\acknowledgments
We are grateful to W.~Kummer for a long-time collaboration on $2D$ dilaton
%NEWDANIEL
gravity, to V.~Frolov, M.~Kreuzer, E.~Scheidegger, T.~Strobl and H.~Verlinde 
for helpful comments, to the anonymous referee for his excellent 
constructive criticism, and to S.~Alexandrov for bringing this problem to our 
attention. This work has been supported by project P-14650-TPH of the Austrian 
Science Foundation (FWF) and the MPI MIS (Leipzig).

\appendix

\section{All dilaton-shift invariant models}\label{app:A}

Our goal is to find the most general action (\ref{eq:dvva1}) with $X=f(\phi)$,
where $f$ is an arbitrary smooth and invertible function, which is invariant
under arbitrary shifts
\eq{
\phi(x)\to\phi(x)+\phi_0\,,\hspace{0.5cm} \phi_0\in\mathbb{R}\,. 
}{dvvshift}
By ``invariant'' we mean that the solutions change by a gauge transformation
so that curvature and torsion do not 
change.

So we try to answer two questions: what is the most
general potential ${\cal V}(X,X^aX_a)$ compatible with dilaton-shift invariance
(\ref{dvvshift}) and what does the corresponding function $f(\phi)$ look like?

The EOMs from (\ref{eq:dvva1}) read
\begin{eqnarray}
 &  & dX+X^{-}e^{+}-X^{+}e^{-}=0\, ,\label{eq:a5} \\
 &  & (d\pm \omega )X^{\pm }\mp \mathcal{V}e^{\pm }=0\, ,\label{eq:a6} \\
 &  & d\omega +\epsilon \frac{\partial \mathcal{V}}{\partial X}=0\, ,\label{eq:a7} \\
 &  & (d\pm \omega )e^{\pm }+\epsilon \, \frac{\partial \mathcal{V}}{\partial X^{\mp }}=0\, .\label{eq:a8} 
\end{eqnarray}

Under the transformation (\ref{dvvshift}) $X$ transforms as 
\eq{
X=f(\phi) \to f(\phi +\phi_0 )\,.
}{dvv:ap1}
It is useful to impose Eddington-Finkelstein gauge $e_0^+=0$, $e_0^-=1$, 
$e_1^+=1$. The components of eq.\ (\ref{eq:a5}) together with invariance of 
the Killing norm (which is now the only non-trivial vielbein component) 
imply immediately 
\eq{
X^\pm\to X^\pm \frac{ f'(\phi +\phi_0)}
{f'(\phi)}\,.
}{dvv:ap2}
The invariance of curvature resp. torsion together with (\ref{eq:a7}) resp.
(\ref{eq:a8}) yields
\eq{
\frac{\partial{\cal V}}{\partial X} \to \frac{\partial{\cal V}}{\partial X}\,,
\hspace{0.5cm}\frac{\partial{\cal V}}{\partial X^\pm} \to 
\frac{\partial{\cal V}}{\partial X^\pm}\,.
}{dvv:ap3} 
This implies, that first 
partial derivatives of ${\cal V}$ have to depend only 
on a dilaton-shift invariant combination of $X^+X^-$ and $X$, i.e. on 
$X^+X^-\tilde g(X)=X^+X^-g(\phi )$ with a new function 
$g(\phi )=\tilde g(f(\phi))$ which has to be determined yet. It is 
clear that only one such independent combination can exist\footnote{There are 
three target space coordinates, so at most three independent combinations 
could exist. However, Lorentz-invariance restricts us to two independent 
combinations (normally $X^+X^-$ and $X$) and dilaton-shift invariance 
restricts us further to a single combination.}. The condition fulfilled by 
$g(\phi )$ reads
\eq{
g(f(\phi)) = \left(\frac{
f'(\phi +\phi_0)}{f'(\phi)}\right)^2
g(\phi +\phi_0 )\,.
}{dvv:ap4} 
When looking at infinitesimal shifts $\phi_0$ one obtains a differential 
equation for $g$, the solution of which is
\eq{
g(\phi)=\frac{c}{{f'(\phi)}^2}\,,\hspace{0.5cm}c\in\mathbb{R}\,.
}{dvv:ap5}
Thus we know (up to a constant which we fix to 1) the dilaton-shift invariant 
combination of the target space coordinates. Therefore, the most general 
consistent ${\cal V}$ must be of the form
\eq{
{\cal V} = l(X) U(Z)\,,\hspace{0.5cm}Z:=\frac{X^+X^-}{{f'(\phi)}^2}\,.
}{dvv:ap6}
The conditions (\ref{dvv:ap3}) determine the new function $l$ up to a
multiplicative constant, which can be absorbed into a redefinition of $U$. 
The result is:
\eq{
l(X(\phi)) \propto f'(\phi)\,. 
}{dvv:ap8}
Obviously, knowledge of $f(\phi)$ determines also ${\cal V}$ uniquely.
Up to now we have not used eq.\ (\ref{eq:a6}). Its invariance provides the last
restriction, which is sufficient to calculate the most general $f(\phi)$
compatible with dilaton-shift invariance. It is more convenient to use the
conservation law (which is a certain linear combination of (\ref{eq:a6}) and
(\ref{eq:a5})) instead of (\ref{eq:a6}):
\eq{
d(X^+X^-)+f'(\phi)U(Z)f'(\phi)d\phi=0\,,
}{dvv:ap9}
which can be brought into the form
\eq{
dZ+\left[U(Z)+2Z\frac{f''(\phi)}{f'(\phi)}\right]d\phi=0\,.
}{dvv:ap10} 
Since $Z$ is unaffected by construction we obtain from the invariance of 
(\ref{dvv:ap10}) that $(\ln f')'$ must be invariant. This yields
\eq{
f'(\phi)=c_0\exp{[-2\al\phi]}\,,\hspace{0.5cm}\al\in\mathbb{R}\,.
}{dvv:ap13}
The multiplicative constant $c_0$ can again be absorbed into $U(Z)$. 
The final 
integration involves a new additive constant which would only lead to a 
surface term $d\om$ in (\ref{eq:dvva1}). Thus, we fix it to zero.
For $\al=0$ we obtain the solution (\ref{dvvXtU}), and for 
$\al\in\mathbb{R}\backslash \{0\}$ we get (\ref{dvvXU}). 
This concludes the proof that these 
are the only dilaton-shift invariant actions of type (\ref{eq:dvva1}).

\section{Approximate solutions to the ESBH}\label{app:B}

Having established the non-existence of an action of type (\ref{eq:dvva1})
reproducing the ESBH it is still possible to construct models which mimic
most of its essential features. This is interesting for two reasons: firstly, 
one might still learn something about string theory in the non-perturbative  
regime. Secondly, models with a potential (\ref{dvvXU}) are not studied 
anywhere else and they are interesting on their own.

The problem of constructing $2D$ dilaton gravity actions admitting classical
solutions with given properties is not new. For example, the dilaton
models for black holes with regular de Sitter interior were considered in
\cite{Trodden:1993dm}.
\subsection{Exact dilaton, approximate line element}\label{app:B.1}

Which features do we require? First of all, dilaton-shift invariance has been 
of fundamental importance in the present context, so we want to keep it by all
means. Secondly, we demand
asymptotic equivalence to the ESBH in the weak- {\em and} strong-coupling 
regions. Thirdly, exactly one Killing-horizon should be present and the 
naively calculated Hawking temperature should be constant (by this we mean its
independence on the value of the Casimir (\ref{dvvCW})). Fourthly, we 
require the existence of a ``dual'' solution which has no Killing-horizon, 
because by applying the momentum/winding mode duality to the ESBH one obtains 
a dual solution with precisely these features. Fifthly, we want the ESBH 
solution (\ref{eq:dvv3}) for the dilaton $\phi$. And finally, our model should 
have the Minkowski ground state property like the ESBH, i.e. there must exist 
one value ${\cal C}_0$ of the Casimir (\ref{dvvCW}) where the 
metric yields Minkowski 
space.

If these requirements can be met we have a model which is very similar to the 
ESBH and where all differences are confined to the line-element -- and even
that must approximate the ESBH line element very well in two different limits.

In fact, any solution of eqs.\ (\ref{eq:dvvnew1})-(\ref{dvvdkdZ}) fulfills
 1.\ and 5.\ by construction. It is defined almost uniquely -- the only 
essential parameter that we have at 
our disposal is $\al$. Point 4.\ restricts us to $\al=\mathbb{N}_{\rm odd}$ 
due to the following observation: in (\ref{dvvyy0}) we have in general branch
cut ambiguities; we would like to have exactly two branches for the two 
``dual'' solutions. Note that this (sign) ambiguity can be reabsorbed into 
$y_0$ -- e.g. for negative values of $y_0$ we have the ``ordinary'' solution 
while for positive values we have the "dual" one. Of course, the (non-)existence 
of horizons still has to be checked. Moreover, as the limit $p\to 0$ in 
(\ref{dvvp1}) proves we need $\al=1$ to obtain the correct weak coupling 
limit. Thus, 2.\ restricts us to $\al=1$ (fortunately compatible with the 
prior restriction). Having fixed our essential
parameter we have yet to check whether the other requirements can be met.

The solution of (\ref{dvvdkdZ}) for $\al=1$ turns out as\footnote{For general 
$\al$ the solution is given by $z\cdot\hyperg(\al/2,-\al/2,1+\al/2;-z^{2/\al})=
(Z-Z_0)/y_0$. For $\al=1$ this belongs to the degenerate class of 
hypergeometric functions, which is why we obtain the simpler form 
(\ref{dvv:apb1}).} 
\eq{
z\sqrt{1+z^2}+\arcsinh{z}=2\frac{Z-Z_0}{\tilde{y}_0}\,,
}{dvv:apb1}
with
\eq{
z:=\frac{k}{\sqrt{B}}\,,\hspace{0.5cm}\tilde{y}_0=y_0B\,.
}{dvv:apb2}
This provides an alternative prove of the non-existence of an action of type 
(\ref{eq:dvva1}) reproducing the ESBH: since (\ref{dvv:apb1}) is 
non-algebraic, but (\ref{dvvZX}) is algebraic they cannot be 
equivalent except 
at certain points. For $y$ in terms of $z$ eq.\ (\ref{dvvyy0}) yields
\eq{
y = \tilde{y}_0 z \sqrt{1+z^2}\,.
}{dvv:apb4}
By virtue of (\ref{eq:dvvnew1}) and (\ref{dvvyy0}) $U$ can be expressed
as a function of $z$:
\eq{
U(Z)=-\tilde{y}_0\,\arcsinh{z(Z)}+2Z_0
}{dvv:apb3}
With the redefinition $\tilde{U}:=2(U-2Z_0)/\tilde{y_0}$ a
non-algebraic equation for $\tilde{U}(Z)$
\eq{
\tilde{U}+\sinh{\tilde{U}}= - 4\frac{Z-Z_0}{\tilde{y}_0}\,,
}{dvv:apb5}
is established.
The solution depends on one integration constant ($Z_0$) and an additional
parameter ($\tilde{y}_0$), both of which depend on $p$. For $p\to 0$
$\tilde{y}_0$ tends to infinity, for $p\to 1$ it vanishes. In these limits we 
recover CGHS and JT, respectively,
\eq{
U(Z;p\to 0) = Z_0 - Z + {\cal O}(p)\,,\hspace{0.5cm} U(Z;p\to 1) = 2Z_0 + 
C(p)\ln{(Z-Z_0)}\,,C(1)=0\,,
}{dvv:apb6}
because ${\cal V}(X,Z;p=0)=-X^+X^-/X+Z_0X$ and ${\cal V}(X,Z;p=1)=2Z_0X$, in 
accordance with our second requirement\footnote{
For small but non-vanishing $p$ we obtain still an algebraic solution for $U$ 
and one only has to solve a cubic equation $z^3/6+z=(Z-Z_0)/\tilde{y}_0$. The 
limit $p\to 1$ is somewhat singular (see below). Thus, unfortunately the strong
coupling region is not fully under control despite of the nice JT limit. Note,
however, a slight discrepancy as compared to eqs.\ (\ref{dvv:VCGHS}),
(\ref{dvv:VJT}): if $Z_0$ is chosen as $-2b^2$ the CGHS is produced correctly,
while JT differs by a factor of $1/4$.}.

Remembering (\ref{eq:dvvnew2}) and (\ref{dvv:apb2}) one can use
(\ref{dvv:apb1}) to express $Z$ as a function of the dilaton $X$. This is of
importance for the third and the fourth requirement, since the Killing-norm
expressed as a function of $X$ is given by
\eq{
\xi(X)\propto \frac{Z(X)}{X^4(Z'(X))^2} \propto \frac{X^2Z(X)}{w^2}\,,
}{dvv:apb7}
with $w$ defined in (\ref{dvvw}) -- note that $w$ is strictly positive and thus
there are no singularities in (\ref{dvv:apb7}). So the question of 
(non-)existence of Killing-horizons reduces to the question of zeros in $Z(X)$.
We need two ingredients: first, observe the strict monotony of the left 
hand side of eq.\ (\ref{dvv:apb1}); second, note the strict 
positivity/negativity of $z$ due to the strict positivity/negativity\footnote{
By virtue of its definition $X$ must be strictly positive, but ${\cal C}_0$ 
can be positive or negative. For ${\cal C}_0=0$ we obtain the (Minkowski) 
ground state, i.e.\ the (trivial) vacuum solution.}
of ${\cal C}_0/X$ in (\ref{eq:dvvnew2}). Thus, depending on the sign of 
$Z_0/\tilde{y}_0$ (which is a fixed constant for each value of $p$) there is a 
Killing horizon or there is none. Suppose that we have fixed all constants 
such that a horizon exists for a given $p$. Then, by changing the sign of $y_0$
(in other words, by choosing the other branch present in the solution of 
(\ref{dvvyy0})) we obtain another model with the same value of $p$ and all 
other constants, except that no Killing-horizon exists for that (``dual'') 
model.

Concerning the issue of Hawking temperature we use its definition in terms of
surface gravity (cf. e.g. \cite{waldgeneral})
\eq{
T_H = \frac{1}{4\pi} \left. \frac{d\xi}{dr} \right|_{r_h}\,,
}{eq:dvv108}
plugging in the relations
\eq{
\frac{d\xi}{dr}=\frac{d\xi}{dX}\frac{dX}{dr}\propto\frac{d\xi}{dX} I^{-1}(X)\,.
}{dvv:apb8}
Since we have to evaluate (\ref{dvv:apb8}) only at the horizon calculations 
simplify considerably: we have to act with $d/dX$ only on $Z(X)$ in 
(\ref{dvv:apb7}) and obtain finally $T_H\propto\rm T_0$, where $T_0$ does not 
depend on the value of the Casimir function (\ref{dvvCW}). Thus, it is a 
universal ($p$-dependent) constant in accordance with our second requirement.

Finally, the Minkowski ground state property must be examined. From 
(\ref{eq:dvv52}) a necessary and sufficient condition for flatness is 
\eq{
\left. \exists \, {\cal C}_0,c_1\in\mathbb{R} \, \right| \,
X^2W^{-1}\left(\ln{\frac{{\cal C}_0}{X}}\right) I^2(X) = c_1\,.
}{eq:dvv64}
There are two possibilities  to satisfy this condition: either, it is satisfied
independently of ${\cal C}_0$ or there exist just certain values (at least 
one) of ${\cal C}_0$ for which the relation (\ref{eq:dvv64}) holds. The first 
case implies
\eq{
\frac{U'_{MGS}}{U_{MGS}}=\frac{1}{2Z}\, \rightarrow \,U_{MGS} = c\sqrt{Z}\,, 
c\in\mathbb{R}\,,
}{eq:dvv65}
which appears to be a rather pathological solution because it means that for 
arbitrary values of the Casimir function one always obtains Minkowski space as 
a solution for the line element but a non-vacuum solution for the dilaton.
This just reinforces the common knowledge that not every toy model one can 
make up needs to make physical sense. Incidentally, the potential $U_{MGS}$ is
the only one satisfying both dilaton-shift invariance conditions (\ref{dvvXtU})
and (\ref{dvvXU}).
In the more interesting case of an isolated solution we assume that the 
quantity $X^2I^2(X)$ becomes a constant\footnote{This assumption is not 
necessary in general, but sufficient for the present case.} for a certain 
value of ${\cal C}_0$. Then, for the same value of ${\cal C}_0$ the function 
$W^{-1}\left(\ln{{\cal C}_0/X}\right)$ must 
be constant. Even under this restricted assumptions we encounter still
two possibilities: either $I(X)=c/X$ for all values of ${\cal C}_0$ and 
$W^{-1}(\ln{{\cal C}_0/X})= {\rm const.}$ for a particular value of 
${\cal C}_0$ -- this is the case for the CGHS model -- or $I(X)=c/X$ only at a 
certain value of ${\cal C}_0$ and simultaneously 
$W^{-1}(\ln{{\cal C}_0/X})= {\rm const.}$ with the {\em same} value of 
${\cal C}_0$. We focus on the latter as it applies to our model. 
This non-trivial case implies the 
existence of a ${\cal C}_0^*\in\mathbb{R}$ such that
$I(X,{\cal C}_0^*) = c_1/X\,,\, c_1\ne 0\,,$
and
$Z(X,{\cal C}_0^*) = c_2\,,\, c_2\ne 0\,,$
thus promoting the Killing norm
$\xi(X,{\cal C}_0^*) = 2c_1^2c_2$
to a constant (which can be adjusted to 1 by rescaling the 
coordinate $u$) and 
hence the line element (\ref{dvvEFg}) describes Minkowski space for that 
particular value of ${\cal C}_0^*$. 

So it was indeed possible to fulfill all required features, which proves that
dilaton gravity with a potential ${\cal V}=XU(Z)$ and $U(Z)$ given in 
(\ref{dvv:apb3}) is very close to the ESBH solution. Since the only 
independent quantity which deviates from the latter is the Killing-norm we can 
quantify this statement in a simple manner. This is done most conveniently 
using $\xi$ as a function of $X$ in the ratio
\eq{
{\cal R}(X;p):=\ln{\left|\frac{\xi_{ESBH}(X;p)}{\xi_{\rm our model}(X;p)}
\right|} = \ln{\left|\frac{1-2/(p(1+w))}{\mu X^2Z/w^2}\right|}\,.
}{dvv:apb10}
For numerical plots we still have to fix the six constants $b,\mu,\phi_0,
{\cal C}_0,Z_0,\tilde{y}_0$. It turns out that once the overall scale is 
chosen\footnote{We fix it such that for the ESBH the Killing-horizon is 
located at $X=1$.} there remain only two independent constants (e.g. $Z_0$ and 
$\tilde{y}_0$). We employ the conditions 1.\ $\lim_{X\to\infty}\xi(X)=1$, 
$\forall p\in(0;1)$ and 2.\ the NLO terms in a $1/X$ expansion should be equal
for both Killing-norms (this corresponds to the proper ADM term). 
This model breaks down for $p$ close to 1 and $X$ close to 0, despite of the 
correct JT limit $p\to 1$ discussed above (in that limit it is not sensible to 
impose the asymptotic condition $\xi\to 1$). Thus, unfortunately, we do not 
really describe the strong coupling region\footnote{Of course, we can 
determine the two relevant constants also by fixing the first two terms in an 
expansion near $X=0$. Then, the strong coupling region will be described well 
and deviations become non-negligible in the weak coupling regime.} very
well and point 2.\ of our requirements is seriously challenged.
\DOUBLEFIGURE{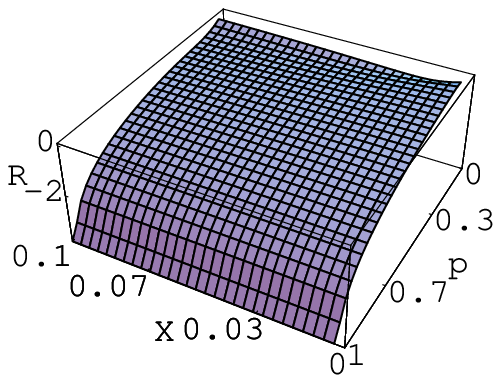}{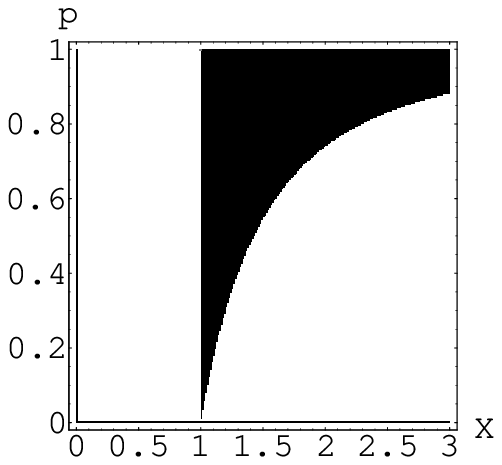}{${\cal R}$ near the origin}{Signum of $\xi_{ESBH}/\xi_{\rm our model}$}
This is clearly shown in the figures. Fig.\ 1 is rather self-explanatory; Fig.\
2 has to be discussed: If the sign of $\xi_{ESBH}$ equals the sign of 
$\xi_{\rm our model}$ the point is depicted in white, else in black. At $X=1$ 
the ESBH Killing-norm vanishes; at a larger ($p$-dependent) value of $X$ our 
model has a Killing-horizon. This means, that our approximate solution shifts 
the horizon a little bit for small $p$ and quite a bit for $p\to 1$.

The reason of incompatibility between 
strong and weak coupling region may be the Minkowski ground state property
which contradicts the fact that all solutions of the JT model have a
non-zero constant curvature. 

Analogous studies can be performed for the ``dual'' solution 
(\ref{dvv:dualkilling}).

\subsection{Exact line element and approximate dilaton?}

By dropping the fifth requirement it could be possible to fix instead the line 
element equivalent to the ESBH. Thus, we automatically would have all its nice 
geometric properties, however at the cost of a deviating solution for the 
dilaton. We do not follow this route explicitly, but even if it 
works\footnote{It is not at all clear that for any given line-element a 
dilaton-shift invariant action (\ref{dvva3}) can be constructed. We have 
neither a prove nor a counter example, but it should be possible to construct
either along the following lines: assume $e^\pm$ as given; take the EOMs
(\ref{dvveo})-(\ref{dvveXpm}) and try to eliminate all variables; 
since there are (in components) nine equations but only six unknown functions 
($\om_\mu,X,X^\pm$ and $U$) either produce a contradiction or extract a 
(unique ?) potential $U(Z)$.}
we expect similar problems: although, for instance, the weak coupling region 
will be described well and the limit $p\to 1$ will yield the JT model the 
strong  coupling region will differ quantitatively from the ESBH 
solution.

\subsection{A ``nice'' potential approximating the ESBH}\label{app:B.3}

The equations (\ref{dvvZX}) and (\ref{dvvUX}) can be used to extract the 
potential $U(Z)$. 
As proved in subsection \ref{sec:4.3} it cannot produce the ESBH. Remarkably, 
this approach still yields a very nice result resembling the ESBH in several 
features, which is why we present it nevertheless: it fulfills all required
properties listed at the beginning of subsection \ref{app:B.1}, except for the 
fifth one; on top of that the function $U(Z)$ is purely algebraic, a mayor 
advantage as compared to eq.\ (\ref{dvv:apb5}). Thus, it may be a suitable 
starting point for a toy model study of the ESBH.

The function $U(Z;p)$ is plotted\footnote{Again some constants have to be fixed
conveniently: $b\stackrel{!}{=}1\stackrel{!}{=}\al$; 
$U(w\to\infty)\stackrel{!}{=}-4$} in Fig.\ 3. 
\EPSFIGURE{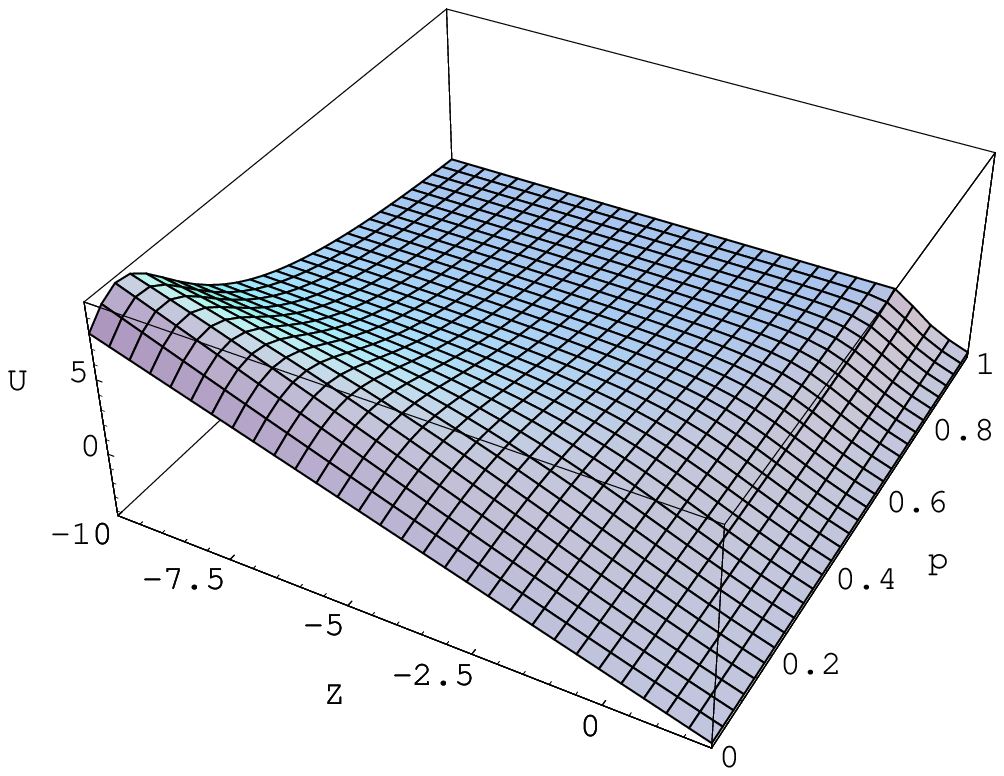}{The ``nice'' potential $U(Z)$} 
The value $Z=2$ corresponds to the asymptotic limit $X\to\infty$, unless $p=1$;
in that case $Z$ stops\footnote{Like in the model discussed in \ref{app:B.1} a 
somewhat mysterious factor of 1/4 is 
involved in the JT limit. Whether this is numerical coincidence or something
deeper has yet to be decided.} at $1/2$. This shows (as probably expected), 
the non-smoothness of the limit $p\to 1$ in the asymptotic region. The horizon 
is located at $Z=0$ and the
``origin'' $X\to 0$ corresponds to $Z\to-\infty$. One sees clearly the linear 
behavior of $U(Z)$ in the CGHS limit $p\to 0$ and the constant behavior 
(until $Z$ reaches the kink point $1/2$) in the JT limit $p\to 1$. So despite
of being the result of a procedure inconsistent with the ESBH, $U(Z)$ has some 
very attractive properties and in principle it can be used as an approximation 
to the ESBH with the correct behavior in the strong and weak coupling regions:
the JT limit is approached as $\lim_{Z\to -\infty} U(Z;p)=3/p-4$; close 
to $Z=2$ the model behaves like the CGHS, 
$U(Z\approx 2;p)=-4-(Z-2)+p(2+p)(Z-2)^2/8+\dots$

%As a further illustration the function $U(Z;p=8/9)$ is printed in Figs.\ 6 and 7.
%\DOUBLEFIGURE{U89origin.eps}{U89.eps}{$U(Z;p=8/9)$ near the ``origin'' ($Z\to-\infty$)}{$U(Z;p=8/9)$ in the horizon- ($Z=0$) and asymptotic region ($Z=2$)}
%Obviously, close to $Z=2$ the model behaves like the CGHS, while for 
%$Z\to-\infty$ it approaches the JT with $\lim_{Z\to -\infty} U(Z;p=8/9)=
%3/p-4=-5/8$.

For sake of completeness the explicit solution is provided as well. We only
have to integrate (\ref{dvvUX}) once and choose the integration constant 
conveniently,
\eq{
U(Z)= -4b^2\left(1-\frac{1}{p(1+w(Z))}-\frac{1}{p(1+w(Z))^2}
\right)\,,
}{ap:b3.1}
where $w(Z)$ is a solution obtained from inverting (\ref{dvvZX}). It is unique
due to the following observations: the discriminant of this cubic 
equation\footnote{Not very surprisingly the cubic equation becomes 
algebraically special at the endpoints $p=0$ and $p=1$: in the former case, it
reduces to a linear equation (yielding the CGHS solution), in the latter to a 
quadratic one (yielding the JT solution).} 
changes its sign at the Killing horizon; for negative $Z$ it is negative and a
unique real solution exists in that region; continuity at the horizon allows a
unique matching to the region of positive $Z$ (and positive discriminant) where
three real solutions exist.

A ``dual'' model is again obtained by replacing $w\to -w$ in (\ref{dvvZX}) 
and (\ref{ap:b3.1}), i.e. by going to the second branch of (\ref{dvvw}).

\bibliographystyle{JHEP}
\bibliography{../review01/review,dvv}

\end{document}